\documentclass[twocolumn,trackchanges]{aastex701}
\usepackage{xcolor}

\newcommand{\R}[1]{\textcolor{black}{#1}}

\begin{document}

\title{High-energy Processes in the Bubbles of Wolf-Rayet Stars:\\
The case of WR\,102}

\author[orcid=0009-0003-7442-9918,gname='Luna',sname='Espinosa']{Luna Espinosa}
\affiliation{Instituto de Astronomia, Geofísica e Ciências Atmosféricas, Universidade de São Paulo}
\email[show]{luna.espinosa@usp.br}  

\author[orcid=0000-0002-5444-0795,gname='Maria Victoria', sname='del Valle']{Maria Victoria del Valle} 
\affiliation{Instituto de Astronomia, Geofísica e Ciências Atmosféricas, Universidade de São Paulo}
\email{mvdelvalle@usp.br}

\begin{abstract}
Supersonic winds from massive stars carry great amounts of kinetic power and modify the surrounding interstellar medium. Through this interaction a stellar bubble is formed. Theoretical studies and recent observations suggest that the winds of massive stars could be sources of Galactic cosmic rays. The first detection of synchrotron emission from the bubble of a single star was reported, indicating the presence of relativistic electrons. 
Studying the non-thermal emission from a single massive star can help to better understand the acceleration of particles taking place in massive star clusters. WR\,102 is the perfect case of study. In this work, we present the first high-energy model for the bubble of WR\,102: G2.4+1.4.  
We aim at fitting the radio data and predicting gamma-ray emission. We assume that both electrons and protons are accelerated at the wind shock. We applied a classical model for the stellar bubble and adopted a one-zone model for estimating the radiation produced by the relativistic particles near the acceleration region. Additionally, we computed the expected emission from the protons that diffuse to the outer regions of the bubble. Also, we estimated the leptonic and hadronic contributions expected from cosmic rays. We fitted the observations considering that 3\% of the wind kinetic power goes into relativistic electrons, and a magnetic field of 250\,$\mu$G. The dominant component at high energies is produced by locally accelerated protons reaching the shell. Protons might reach PeV energies in the wind bubble, but the predicted gamma-ray flux is too low to be detectable.


\end{abstract}



\section{Introduction}

During the last decade, massive stars have gained significant attention as sources of Galactic cosmic rays and gamma rays. There are both theoretical and observational reasons for this \citep[see, e.g.,][and references therein]{2019NatAs...3..561A,2025BAAA...66...73D}. However, the strongest one is the recent high- and very-high-energy ($E>100\,$TeV) emission observations related to massive star clusters \citep[e.g.,][]{2007A&A...467.1075A,2011Sci...334.1103A,2021NatAs...5..465A,2022A&A...666A.124A,2024ApJS..271...25C}. \R{Still}, the idea that massive stars can accelerate particles in their winds is not new \citep[e.g.,][]{1980ApJ...237..236C,1982ApJ...253..188V}.  


Massive stars modify their surroundings through their very intense radiation fields, ionizing a large region of parsecs size around them. Also, they have very powerful winds, \R{with velocities over 1000 km/s}, that inject great amounts of kinetic power into the medium and shock it. The stellar wind carves a stellar bubble around it. The combined effect of stellar winds in massive clusters produces a \emph{superbubble} \citep{1977A&A....59..161D, Weaver1977, 2008IAUS..250..341C}. Also, through the evolutionary stages that stars transit, mass-loss episodes enrich the interstellar medium (ISM) with heavy elements. 

In the interaction of the stellar wind, that has velocities of \R{thousands} of kilometers per second in massive stars, with the ambient medium, a system of two shocks is formed. One shock develops in the ISM (often called the \emph{forward} shock), and the other propagates through the wind (often called the \emph{reverse} or \R{termination} shock). The forward shock is a slow shock and loses much of its energy through radiative cooling (i.e., a radiative shock). The reverse shock, on the other hand, is fast and strong (i.e., an adiabatic shock), reaching speeds of the order of the wind velocity. See \citet{2023IAUS..370..205M}, and references therein. 

In the early 2010s, the bowshock of a runaway massive star was discovered to emit non-thermal radio emission \citep{2010A&A...517L..10B}. This observation confirms that massive stars have the capability to accelerate, at least electrons, up to relativistic energies. In 2012, \citet{delvalle2012} presented the first model for gamma-ray emission of bowshocks of runaway stars, soon after many other models followed, together with other candidates \citep[e.g.,][]{2018A&A...617A..13D, 2018ApJ...861...32S,2022A&A...663A..80M,2022MNRAS.512.5374V}. \citet{delvalle2018} concluded  that given the better sensibility of current instruments at radio wavelengths, these systems are more prone to be detected at radio through the synchrotron emission they produce than at gamma energies.


The acceleration of particles responsible for the emission can have 3 different origins: fresh acceleration of particles in the reverse shock \citep[e.g.,][]{delvalle2012}, fresh acceleration of particles in the forward shock, reacceleration and compression of cosmic rays in the forward shock \citep[e.g.,][]{2022A&A...663A..80M,2024MNRAS.530..539M}. The interaction of collective stellar winds in star clusters forms a more complex structure, and the mechanism behind the acceleration of relativistic particles is not well understood.

Extended gamma-ray emission from the bubble of a massive star was reported by \citet{2018A&A...615A..82M}. The authors claim that the emission is associated with $\kappa$\,Ori, a blue supergiant B0.5 Ia driving an intense wind, located at $d = 200\,$pc. The excess of gamma-ray flux, center in this star, was found in data from the {\it AGILE} gamma-ray satellite. The excess location is compatible with the position of the wind termination shock.
The high-energy radiation might have been produced by the interaction of relativistic particles accelerated \R{or re-accelerated} in this fast shock \R{\citep{2019A&A...622A..57C}}. If confirmed, this is the first direct detection of gamma-ray emission from a single stellar wind bubble. 

The first bubble of a single massive star to be reported as a radio non-thermal emitter is G2.4+1.4 \citep{Prajapati2019}. This bubble is produced by the very energetic Wolf-Rayet (WR) star WR\,102, located at 2.88\,kpc \citep{2019A&A...621A..92S}. The authors reported a negative spectral index at low frequencies, compatible with synchrotron emission. 

\R{More recently, diffuse radio emission with a negative spectral index was reported from the WR bubble NGC\,2359, associated with WR\,7 \citep{2026arXiv260202923S}. This represents only the second detection of this kind and further supports the idea that wind-driven shocks in isolated WR bubbles can accelerate particles to relativistic energies.}


Wolf-Rayet stars are highly interesting sources at high energies for multiple reasons. For example, they are \R{likely} progenitors of core collapse supernovae, \R{in particular SN Ibc \citep[e.g.,][]{2012A&A...544L..11Y}}.\R{WR stars are classified based on the strength of different emission lines. Among them, oxygen-rich WR stars (WO) constitute a rare and short-lived evolutionary phase of massive stars. The main indicator of this subtype is the presence of a strong O \textsc{VI} $\lambda$3811-34\,{\AA} emission line. Other major subtypes are characterized by nitrogen (WN) and carbon (WC) dominant emission features.} Also, WC and WO are black hole progenitors \citep[e.g.,][]{2015A&A...581A.110T}. Additionally, the winds of WR have been proposed to account for most of the isotopic anomalies observed in cosmic rays \citep[e.g.,][]{1982ApJ...258..860C}. Furthermore, a small number of WR stars can equal or even outweigh the influence of a whole population of nearby OB stars \citep[e.g.,][]{2018A&A...615A..40R}. The latter is of special interest when studying massive star clusters as gamma-ray emitters.

In this work, we present the first high-energy model for the bubble of WR\,102: G2.4+1.4. We aim at fitting the radio data and predicting gamma-ray emission from the system. Studying the non-thermal emission from a single massive star can help to better understand the acceleration of particles taking place in massive star clusters. WR\,102 is the perfect case of study.  

This work is organized as follows. In the next Section, we briefly introduce the system under study and the reported non-thermal observations. In Section\,\ref{sec:model}, we describe our model and we present our results in Section\,\ref{sec:results}. In Section\,\ref{sec:discussion}, we discuss the implications of our results, and finally in Section\,\ref{sec:conclusions}, we present a summary and give our conclusions.    

\section{The bubble G2.4+1.4}\label{sec:system}

The star WR\,102 was first detected by \citet{1968ApJ...152.1015B}, and later classified as an oxygen-rich Wolf-Rayet star of subtype WO2. These subtypes of Wolf-Rayet stars 
reflect the last observable stage before core collapse. WR\,102  is believed to be in an advanced stage of evolution, it is very hot with an effective temperature $T_{\rm eff} = 200$\,kK. WR\,102 drives a very powerful wind of $V_{\rm w} = 5000$\,km\,s$^{-1}$ and mass-loss rate of $\dot{M} = 5.8\times10^{-6}\,$M$_{\odot}$yr$^{-1}$ \citep{Sander2019}. 

WR\,102 feeds a wind bubble cataloged as G2.4+1.4 \citep{Dopita1990}. The bubble exhibits an inhomogeneous appearance, with dense filaments and ring-like structures, most probably due to successive interactions of previously ejected material. This asymmetric bubble is observed in multiple wavelengths: at radio, infrared and H$\alpha$ \citep[e.g.,][]{1995ApJ...439..637G,2015A&A...578A..66T,2025RNAAS...9..196Q}.     

The non-thermal emission from the bubble was obtained with the Giant Meterwave Radio Telescope (uGMRT), in the bands 4 (550-850\,MHz) and 5 (1050-1450\,MHz). The continuum maps reveal the characteristic morphology of G2.4+1.4, observed at other frequencies \citep{Prajapati2019}. The computed integrated flux density varies between 2.6 to 1.2\,Jy in the frequency range 605-1429\,MHz. From the flux density measured $S(\nu)$, the authors reported a global spectral index of \R{$\alpha_{\rm sync} = 0.83 \pm 0.10$, this is $S(\nu) \propto \nu^{-\alpha_{\rm sync}}$. }

\section{The model}\label{sec:model}
\subsection{The bubble}

\begin{figure}
\includegraphics[width=\columnwidth]{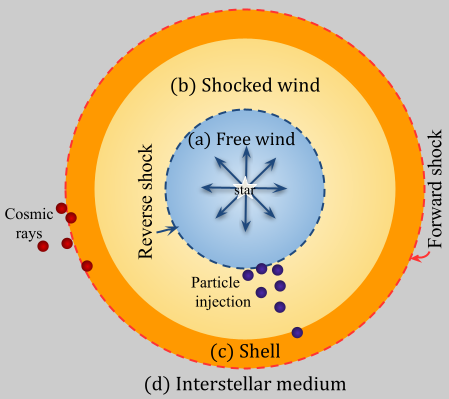}
\caption{Scheme of a typical stellar \R{wind} bubble following the model by  \citet{Weaver1977}, not to scale. Also shown are the location sites of particle injection and background cosmic rays.
\label{fig:bubble}}
\end{figure}


We adopted the classical model for a stellar bubble by \citet{Weaver1977} to model the bubble structure. The interaction of the wind with the ambient ISM produces a stratified spherical structure divided into four distinct regions. These are: the free-flowing wind (region $a$), the shocked wind (region $b$), the shocked interstellar medium (region $c$), and finally the unperturbed medium (region $d$). Regions $a$ and $b$ are separated by the reverse shock, while the regions $c$ and $d$ are separated by the forward shock. A sketch of the bubble is shown in Fig.\,\ref{fig:bubble}, \R{representing the intermediate stage of the stellar wind bubble, also called ``snowplow''. At this stage, thermal conduction from region $c$ to region $b$ significantly impacts the structure of the bubble.}

The reverse shock is very fast, with velocities of the order of the wind velocity, this is thousands of kilometers per second. The forward shock speed depends on time and, \R{differentiating the integrated similarity solution for the radius given by} \citet{Weaver1977}, can be written as:
\begin{equation}
v_{\rm r}(t)   = \frac{3}{5}\,\beta \left(\frac{P_{\rm w}}{\rho_{\rm ISM}}\right)^{1/5} t^{-2/5},
\end{equation}
here $\rho_{\rm ISM}$ is the mass density of the unperturbed interstellar medium and  $P_{\rm w}$ is the wind power. The coefficient $\beta$ depends on the evolutionary stage of the bubble; through this work we consider an intermediate stage for G2.4+1.4, corresponding to $\beta$ = 0.76 \citep{Weaver1977}. 

The radius at which the forward shock is located, $R_{\rm{f}}$,  can be considered as the radius of the stellar bubble as measured from the observations. We adopted  $R_{\rm{f}} \sim 2.51$\,pc from \citet{Prajapati2019}. This value is also in agreement with numerical models of G2.4+1.4 \citep{Brighenti1995}\R{, although the exact geometry of the region does not substantially alter the results \citep[see][]{2018A&A...620A.144D}.} For consistency, we derived the age of the system from the expression of $R_{\rm{f}}$ predicted by the model \citep{Weaver1977}:
\begin{equation}
    t_{\rm{age}} = R_{\rm{f}}^{5/3}  \left ( \frac{250}{308\pi} \right)^{-1/3}P_{\rm{w}}^{-1/3}\rho_{\rm{ISM}}^{1/3}.
\end{equation}
This result is in accordance with the characteristic dynamical time-scale of WR nebulae \R{\citep[e.g.,][]{Brighenti1995, 2014A&A...564A..30G}.} 

The position of the reverse shock, $R_{\rm{r}}$, can be computed from \citep{Weaver1977}:

\begin{equation}
    R_{\rm{r}} = 0.9\,\beta^{3/2} \left ( \frac{\dot{M_{\rm{w}}} }{\rho_{\rm{ISM}}}\right )^{3/10}V_{\rm{w}}^{1/10}\,t_{\rm{age}}^{2/5}.
\end{equation}

Adopting a unique value for $\rho_{\rm{ISM}}$ is not easy, given that the region surrounding the bubble, and the bubble itself, presents strong density gradients. The density ranges from  2--4 cm$^{-3}$, up to 40--60 cm$^{-3}$ \citep[e.g.,][]{Dopita1990}. We used an intermediate value of $n_{\rm{ISM}} \sim 30$\,cm$^{-3}$. 

The parameters describing the star and the bubble are shown in Table\,\ref{tab:parameters}. The parameters derived from observations are separated from those derived from the model by a horizontal line.

\setlength{\tabcolsep}{1.5em}
\begin{table}[]
\centering
\begin{tabular}{cc}
\hline
Parameter & Value
\\ \hline
$R_{\star}$ ($R_\odot$) & 0.52
\\
$\dot{M}_{\rm{w}}$ ($M_{\odot}$\,yr$^{-1}$) 
& $5.4\times10^{-6}$
\\
$V_{\rm{w}}$ (km s$^{-1}$)
& 5000
\\
$P_{\rm{w}}$ (erg s$^{-1}$)
& $4.2\times10^{37}$
\\
$T_{\star}$ (kK)
& 200
\\
$d$ (kpc) & 2.88
\\
$n_{\rm ISM}$ (cm$^{-3}$) & 30
\\
\hline
$R_{\rm{f}}$ (pc)
& 2.51
\\
$t_{\rm{age}}$ (yr) 
& $1.4\times10^{4}$
\\
$R_{\rm{r}}$ (pc)
& 0.53
\\ \hline
\end{tabular}
\caption{Parameters describing the star, the wind and the bubble. The values derived from observations are separated from those derived from the model by a horizontal line. See the text for further details.}
\label{tab:parameters}
\end{table}


\subsection{The high-energy particles}

Firstly, we fitted the non-thermal radio emission from G2.4+1.4.  We adopted a steady-state scenario, as the expansion of the bubble is very slow. We assumed that the electrons responsible for the observed radio emission are accelerated at the reverse shock, via diffusive shock acceleration (DSA, see Sec.\,\ref{sec:discussion} for a discussion). This implies that the reverse shock is powering the relativistic particles. We followed a similar treatment as that presented by \citet{Prajapati2019}. 

Electrons are injected following a power-law in energy plus an exponential cut-off: $Q(E) = Q_0^e E^{-\alpha} e^{-E/E^{e}_{\rm emax}}$, where $\alpha$ is directly derived from the radio observations, assuming that the mechanism behind the emission is synchrotron, then $\alpha = 2\times \alpha_{\rm sync} + 1 = 2.66$). $Q_0^e$ is the normalization, related to the power in relativistic electrons. We estimated the maximum energy balancing the energy gains by DSA and radiative losses. Electrons lose energy by synchrotron radiation but also by adiabatic losses in the expanding post-shock bubble. We calculated this last as  $t_{\rm{adi}} \sim 4(R_{\rm{f}}-R_{\rm{r}})/V_{\rm{w}}$. The acceleration time is calculated as 
\begin{equation}
    t_{\rm{acc}} = \eta_{\rm{acc}}\frac{E}{\sqrt{2/3}B\,c\,q}, \label{tacc}
\end{equation}
where $\eta_{\rm{acc}} = 2\pi(c/V_{\rm{w}})^2$ is the acceleration efficiency in the Bohm diffusion regime.

The distribution of electrons was calculated by solving the following equation \citep{1964ocr..book.....G}:

\begin{equation}
    \frac{\partial }{\partial E}\left [ \left. \frac{dE}{d t} \right|_{\rm loss} N(E) \right ] + \frac{N(E)}{t_{\rm esc}} = Q(E),
\end{equation}
were $t_{\rm esc}$ is the escape time of the particles, that are advected downstream the shocked wind.

The injected electrons would lose most of their energy near the reverse shock, which allows us to consider, as a first estimative, a one-zone model for computing the leptonic emission.

We calculated the synchrotron spectrum following the approach in \citet{1964ocr..book.....G}
\begin{equation}
    L_{\rm{\gamma}} = \frac{2.62\,e^3B}{hmc^2} \int_{E_{\rm{min}}}^{E_{0}} dE\, N(E) \frac{E_{\rm{\gamma}}}{E_{\rm{c}}}^{1/3}e^{-E_{\rm{\gamma}}/E_{\rm{c}}},
\end{equation}
where $B$ is the magnetic field in the emission region, $E_{{\rm{\gamma}}}$ is the photon energy and $E_{\rm{c}}$ is the characteristic energy.

A fraction $\chi_{\rm{NT}}$ of the kinetic power from the stellar wind would go to relativistic particles, that is, $P_{\rm{NT}} = \chi_{\rm{NT}}P_{\rm{w}}$. $P_{\rm{NT}}$ is related to the normalization constant $Q_0^e$ through the relation $P_{\rm{NT}} = \int_{E_{\rm{min}}}^{E_{\rm{max}}} E\,Q(E)\,dE$, where $E_{\rm{min}} = m_ec^2$. Comparing the computed flux with the observed one, we obtained the values for the two free parameters of our model: $\chi_{\rm{NT}} = 0.028$ and the magnetic field strength $B = 250\,\mu$G. It is worth mentioning that this solution is not unique, different combinations of $\chi_{\rm{NT}}$ and $B$ are possible \citep[see][]{Prajapati2019}. The lowest values of $B$ imply a higher efficiency of the shock in transferring kinetic energy to relativistic electrons; given the limited knowledge that exists on this parameter, we favor here a model with a conservative shock efficiency and invoke magnetic field amplification (see the discussion in Sec.\,\ref{sec:discussion}).

\subsection{High-energy protons}

We considered that the reverse shock, which accelerates electrons, is also capable of accelerating protons. The case of protons is different from that of electrons. Protons do not radiate efficiently near the reverse shock, they  lose energy due to adiabatic expansion during the acceleration. However, the velocity of the material from region $b$ to region $c$ drastically decreases. Immediately behind the shock, the flow velocity is given by the usual Rankine-Hugoniot relation for a strong shock $\sim V_{\rm w}/4$. When reaching the region $c$, at $R_{\rm f}$, the expansion velocity is given by $V(t) = 16\,(n_{\rm ISM}/1\,{\rm cm}^{-3})^{-1/5}\,(P_{\rm w}/10^{36}{\rm erg\,s}^{-1})^{1/5}(t/10^6\,{\rm yr})^{-2/5}\,$ km\,s$^{-1}$ \citep{Weaver1977}. In our case, this gives 35\,km\,s$^{-1}$. The protons then diffuse and reach the dense shell, where radiative losses by proton-proton inelastic collision are important. 

\begin{figure}
\includegraphics[width=\columnwidth]{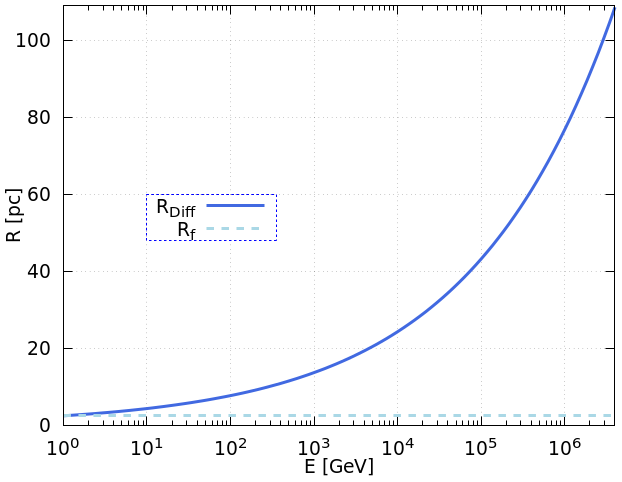}
\caption{Diffusion radius for $t = t_{\rm age}$, as a function of energy.
\label{fig:diff}}
\end{figure}

The injected protons would diffuse through the region $b$; we adopt the diffusion coefficient of the ISM, $D(E) = D_0(E/10\,{\rm GeV})^{0.5}$, but slightly lower, with $D_0 = 10^{26}\,$cm$^{2}$\,s$^{-1}$. For computing the distribution of protons at the shell position ($R \sim R_{\rm f}$) we follow \citet{atoyan1996}. For a continuous injection and under the condition $R_{\rm diff} = 2\times\sqrt{D(E)t_{\rm age}} >> R$, the distribution of protons is \citep{atoyan1996}:
\begin{equation}
    N(E) = \frac{Q_{0}^{p} E^{-2}}{4\pi\,D(E)R}.
\label{eq:protons}
\end{equation}
Here, $Q_{0}^{p}$ is the normalization analogous to that of the electrons already described. For protons we assume $\chi_{\rm NT}^p = 0.1$\R{, based on typical values inferred for supernova remnants \citep[e.g.,][]{2019NatAs...3..561A} and other non-relativistic fast shocks}. We consider that protons are injected with a canonical power-law index of 2. The values of $R_{\rm diff}$ are shown in Figure\,\ref{fig:diff}, for energies $E > 10\,$GeV the above condition is fulfilled. We assume that the shell has a width of $\Delta R \sim 0.5\,$pc.

The maximum energy for the protons is obtained equating $t_{\rm{adi}}$ with the acceleration time \R{(Eq. \ref{tacc})}, that gives $E_{\rm{max\_p}} \sim 4$\,PeV. This value seems rather high. We also consider a different approach to estimate the maximum energy. The acceleration occurs in a region of the size of the shock precursor given by  $D(E)/V_{\rm w}$, here $D(E) = \xi D_{\rm Bohm}$, with $\xi> 1$. Physically, the acceleration takes place in a region of size $l_{\rm p} = R_{\rm r}$, the maximum energy can not exceed $E = 3q\xi^{-1}BR_{\rm r}(V_{\rm w}/c) = 5.9\,$PeV. Taking 10\% of this value gives $E_{\rm max\_p} \sim 600\,$TeV. We further discuss this in Sec.\,\ref{sec:discussion}.

\subsection{Gamma-ray emission}

The high-energy electrons additionally cool by other radiation mechanisms, producing gamma-rays. The electrons suffer inverse Compton scattering against the stellar radiation field (at a distance $R_{\rm r}$) and the cosmic microwave background (CMB).

The losses by relativistic Bremsstrahlung can also be relevant. The shocked wind region of the bubble would be contaminated by material \emph{evaporated} from region $c$ to region $b$, the density in this region can be estimated as  ${n_{\rm b} = 0.01\, n_\mathrm{ISM}^{19/35} (\dot{M}_6 V_{2000}^2)^{6/35}\,t_6^{-22/35}} \text{ cm}^{-3}$ \citep[see,][]{Castor1975}, in our case $n_{\rm b} = 1.6\,$cm$^{-3}$. As mentioned before, the protons collide with the shell material of density $n_{\rm shell}$ at $R\sim R_{\rm f}$ (see Eq.\,(\ref{eq:protons})), producing gamma-rays through neutral pion decay. The shell is formed by cooled interstellar medium, we consider a compress factor of $S = n_{\rm shell}/n_{\rm ISM} \sim 10$, from \citet{Brighenti1995}, then $n_{\rm shell} \sim 300\,$cm$^{-3}$. \R{This value of $S$ is consistent with the expected compression in an isothermal shock $\sim M^2$, with $M$ the shock Mach number. }

The high-energy leptonic and hadronic contributions were calculated using the \textsc{python} package \textsc{naima} \citep{Zabalza2015}. 

\subsection{Thermal radio emission}


Thermal Bremsstrahlung (free-free emission) is expected from this region, and is observed in stellar bubbles \citep{1986ApJ...303..239A}. This thermal component competes with the non-thermal emission at radio frequencies and should be considered in the modeling \citep[e.g.,][]{delvalle2025}.


We estimated the ionized region by calculating the \emph{Strömgren radius}, $R_{\rm st}$, of WR 102 assuming that the stellar radiation field is a blackbody of temperature $T_{\star}$ \citep[e.g.,][]{Maciel2013}.
We obtained $R_{\rm{st}} \sim 8$\,pc, placing the entire bubble within the photoionized sphere. The non-thermal emission in the system under study is stronger than the thermal emission at the observation frequencies. The thermal emission inside the region of flux extraction reported in \citet{Prajapati2019} (of radius $R_{\rm f}$) should be smaller than the non-thermal at the observed frequencies. This helps constrain  the value of the filling factor $\eta_{\rm vol}$ and maintain the consistency of the model. For an ellipsoid of depth $R_{\rm st}$ in the direction of the line of sight, of volume $V =(4/3)\pi R_{\rm f}^2R_{\rm st}$,  we get $\eta_{\rm vol} \sim 1/2$.

\subsection{Gamma-ray absorption}

The stellar photon field can be a source of opacity for gamma rays \citep[e.g.,][]{romero2011}.
We computed the optical depth as \citep{1967PhRv..155.1404G}:

\begin{equation}
    \tau_{\gamma\gamma} = \sigma_{\gamma\gamma} \int_{R_{\rm{r}}}^{R_{\rm{f}}} n_{\rm{ph}}(r) \rm{d}r,
\end{equation}
where $\sigma_{\gamma\gamma}$ is the total cross section for the production of photon-photon pairs and $n_{\rm{ph}}$ is the density of the stellar photon field (assumed monochromatic, with $E_{\rm ph} = k_{\rm B}T_{\star}$), which decreases as $R^{-2}$. In the region of interest, we found that the absorption is negligible.

\subsection{The cosmic-ray contribution}
In extended sources, like the case of wind bubbles, the cosmic rays present in the \R{ISM} can contribute to the total non-thermal emission (see also Sec.\ref{sec:discussion}). Here we computed the contribution from electrons and protons interacting in the dense shell. We used the cosmic-ray flux parametrized in \citet{2015ApJ...810..141P}. The particles suffer compression by the forward shock, modifying their spectra in two ways \citep[e.g.][]{2010ApJ...723L.122U}: increasing the flux by a factor $\Sigma^{2/3}$, with $\Sigma = S/4$, and shifting the distribution as $p \rightarrow \Sigma^{-1/3}p$, being $p$ the momentum of the particles. The magnetic field in the medium was taken as $B_{\rm ISM} = 1\,\mu$\,G, so $B_{\rm shell} = \sqrt{2(S^2-1)/3+1}B_{\rm ISM}$. 

\section{Results}\label{sec:results}

\begin{figure*}[t!]
\resizebox{\hsize}{!}{\includegraphics[clip=true]{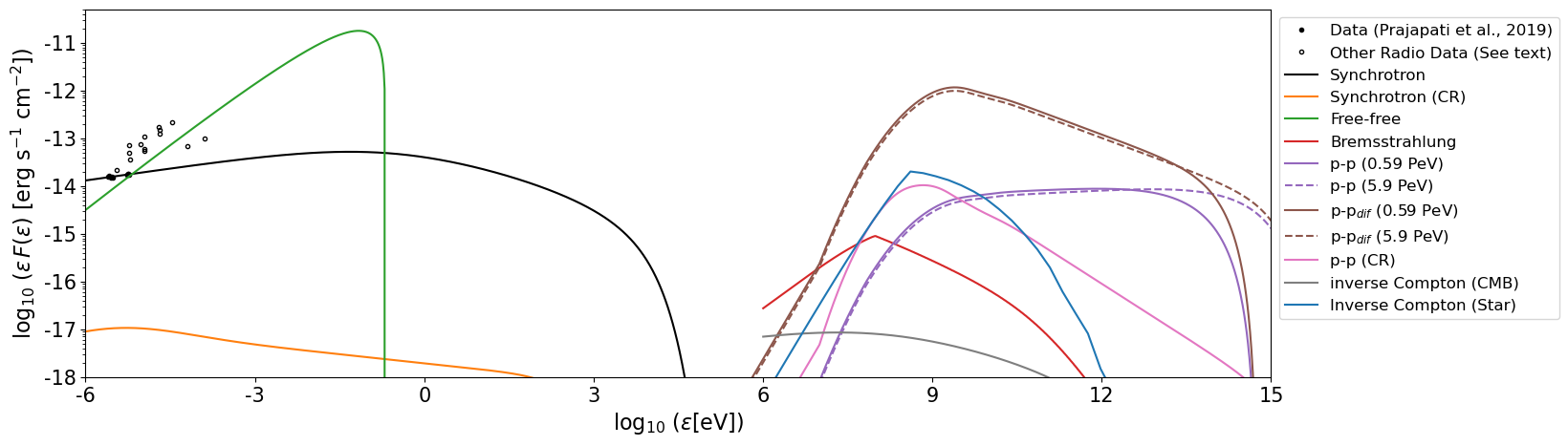}}
\caption{Computed multi-wavelength spectral energy distribution. \R{Observed data are shown as black dots; radio data from \citet{Prajapati2019} are represented by filled black dots, while other radio data from the literature \citet{2022MNRAS.516.3773G} are shown as open circles.} The synchrotron contribution is shown in black. The contributions from free-free emission (green), relativistic Bremsstrahlung (red) and inverse Compton (from the CMB in gray and from the star in light blue) are also shown. The emission from protons at the reverse shock is shown in purple and at the shell in brown, for two maximum energies: 0.6\,PeV (solid) and 6\,PeV (dashed). The synchrotron emission from cosmic-ray electrons (orange) and $p-p$ from protons (pink) interacting in the shell are also shown. 
}
\label{fig:spec-multi}
\end{figure*}

The computed spectral energy distribution (SED) for the bubble is presented in Fig.\,\ref{fig:spec-multi}. \R{The radio data are represented as black points; from \citet{Prajapati2019} (filled dots) and other radio data from the literature \citep{2022MNRAS.516.3773G} (open circles). Although the datasets are not directly comparable (see \citet{Prajapati2019}), we include the data points for completeness.}  The synchrotron contribution is shown in black. The contributions from free-free emission (green), relativistic Bremsstrahlung (red) and inverse Compton (from the CMB in gray and from the star in light blue) are also shown. The emission from protons at the reverse shock is shown in purple and at the shell in brown, for two maximum energies: 0.6\,PeV (solid) and 6\,PeV (dashed). The synchrotron emission from cosmic-ray electrons (orange) and $p-p$ from protons (pink) interacting in the shell are also shown.

At radio frequencies, a competition between thermal Bremsstrahlung and synchrotron radiation occurs, with a non-thermal predominance in the frequencies of detection, which is consistent with the observations from \citet{Prajapati2019}. Our results are also consistent with \citet{1995ApJ...439..637G}. They observed the bubble with VLA (Very Large Array) finding a flat spectrum indicating thermal emission. These observations correspond to a slightly higher frequency, $\nu = 1.49$\,GHz where our model predicts a transition from non-thermal to thermal emission (see Figure\,\ref{fig:spec-multi}).  \citet{1994MNRAS.270..835G} also presented observations of G2.4+1.4 at 843\,MHz; they derived a nearly flat spectral index between 843 and 2695\,MHz. \R{The model is compared with these datasets, in addition to others reported in the literature (see \citet{2022MNRAS.516.3773G}, Table 1). Our results agree with the data and highlight a competition between thermal and non-thermal radiation at the observed bands (see discussion in Sec. \ref{sec:discussion}).}

At high energies, we predict a gamma-ray component with a maximum at the GeV region, produced by the protons accelerated at the reverse shock and interacting at the shell. The produced flux is $\sim 10^{-12}$\,erg\,s$^{-1}$\,cm$^{-2}$. The region of the reverse shock is dominated by the leptonic contribution below 10\,GeV with a maximum of $2.5\times10^{-14}$\,erg\,s$^{-1}$\,cm$^{-2}$, at higher energies the hadronic contribution dominates with $10^{-14}$\,erg\,s$^{-1}$\,cm$^{-2}$.

The radiation from the cosmic rays is not important at the lowest energies. In the case of protons compressed in the shell, the maximum flux produced is $\sim 10^{-14}$\,erg\,s$^{-1}$\,cm$^{-2}$ at $E_{\gamma} \sim 1\,$GeV. However, at those energies the contribution from the electrons interacting with the stellar photon field  exceeds it by a factor of two.

\section{Discussion}\label{sec:discussion}

The magnetic field strength required to fit the observations is of the order of 0.25\,mG. There are no measurements of the magnetic field in WR\,102, however \citet{2014ApJ...781...73D} searched for magnetic fields in 11 bright WR stars using polarization measurements. They reported an average field strength upper limit of $B^{\rm max}_{\rm wind} \sim 500$\,G. We can estimate the magnetic field at the reverse shock position following \citet{1982ApJ...253..188V}. Assuming that the star rotates at a velocity $0.1V_{\rm w}$\footnote{The round emission-line profiles observed in WR\,102 indicates that is a rapid rotator \citep{Sander2012}.}, we have: $B_{\rm r} \sim 0.1\,B^{\rm max}_{\rm wind}({R_\star}/{R_{\rm r}})\left[ 1+ 100(R_\star/R_{\rm r})^2\right]^{1/2}$ giving $\sim 1.1\mu\,$G. Higher values of $B_{\rm r}$ require a magnetic field amplification process.  

A connection between particle acceleration and magnetic field amplification in astrophysical shocks seems to exist beyond shock compression. In non-relativistic shocks, magnetic field amplification factors of $10-20$ can be achieved by non-resonant cosmic-ray instability at the smallest scales \citep[e.g.][]{2014ApJ...794...46C,2018MNRAS.473.3394V}. Additionally, the magnetic field can achieve further amplification, of a factor of 10, at larger scales by a turbulent dynamo powered by the pressure of the accelerated cosmic rays \citetext{e.g., \citealp{2016MNRAS.458.1645D}, Koshikumo et al., \textit{in preparation}}. 

Another important parameter we constrained from the fitting is the fraction of wind power that goes into relativistic electrons. We obtained $\sim 3\%$, which is a conservative quantity. The available power to accelerate particles depends on the mass-loss rate  of the wind. The winds of high-mass stars can be clumpy \citep[e.g.][]{2006ApJ...648..565O}, this could overestimate the prediction of $\dot{M}_{\rm w}$. Even when current models take into account wind inhomogeneities, estimates can still suffer from uncertainties. However, given that only a small fraction of the power is needed to explain the radio emission, this might not be an issue. The dependence of the energy budget on the wind velocity is stronger; also the adiabatic loses and the acceleration efficiency depend on this parameter. Changes in the wind velocity are more difficult to predict {\it a priori}.

Our model predicts a very high value for the maximum energy for protons, of the order of PeVs, under canonical assumptions (see Eq.\,\ref{tacc}), this is under Bohm diffusion. These high values are the result of a combination of the very fast wind of WR\,102 and the magnetic field strength required to fit the observations (see discussion above). The prediction that the wind of a WR star is a PeVatron is highly exciting. As mentioned before, clusters of massive stars are associated with very-high energy sources, making them strong candidates as Galactic PeVatrons. Within star clusters, Wolf-Rayet stars contribute with a large fraction of the collective kinetic power from the stellar winds. Therefore, studying WR stars could be key to understand the capacity of these systems to produce high-energy particles. Detailed models of particle acceleration are needed to better constrain the maximum energy attainable by the protons. 

The emission in the region around 100\,TeV is low, because the protons have to diffuse away from the acceleration region to radiate efficiently. The propagation of the protons naturally softens their distribution spectra.  If G2.4+1.4 is a PeVatron, it would be difficult to confirm it observationally. This could be a limitation faced by most bubbles of massive stars.

There are 705\footnote{\url{http://pacrowther.staff.shef.ac.uk/WRcat/}} WR stars cataloged (up to August 2025), however,  
estimations of the population of WR stars in the Milky Way are between 1900 and 6000 \citep{2009AJ....138..402S,2015MNRAS.447.2322R}. Assuming that all stars have a mechanical wind power of $10^{36}\,$erg\,s$^{-1}$ we can estimate their total contribution to the Galactic cosmic-ray population in protons. If the stars contribute with 10\% of the power in high-energy particles, assuming 6000 individuals, this gives  $\sim$ $6\times10^{39}\,$erg\,s$^{-1}$, 15\% of the total power in cosmic rays in the Galaxy (we take $W_{\rm CR} \sim 4.1\times10^{40}$\,erg\,s$^{-1}$). In the case where the fraction of power in particles is the same as that found for electrons, this contribution is $\sim$ 4.5\%. The contribution from WR stars to the cosmic rays is also important concerning its composition, as their winds are enriched with respect to solar abundances. Of special interest is  $^{22}$Ne, as already mentioned.

\R{Our model predicts a non-thermal flux of $\sim 10^{-15}$\,erg\,cm$^{-2}$\,s$^{-1}$ in the X-ray band, above \textit{XMM-Newton} sensitivity \citep{2001A&A...365L...1J}. However, its capabilities do not allow separation of thermal and non-thermal emission in structures formed by winds of massive stars as found in \citet{2017MNRAS.471.4452D}. Currently, the X-ray band is not an important channel for non-thermal detection in these systems. This emission might be important for the next X-ray facilities, such as the NewAthena mission \citep{2025NatAs...9...36C}.}

The competition between the free-free emission and the synchrotron in the SED is highly dependent on the frequency (see Fig.\,\ref{fig:spec-multi}). The thermal emission grows towards higher frequencies, after equating the non-thermal component, it surpassed the latter by orders of magnitude. The synchrotron emission could only dominate the spectrum at lower frequencies, below $\sim 2.4$\,GHz (see Sec.\,\ref{sec:results}). 

\citet{2022MNRAS.516.3773G} analyzed the integrated radio flux from G2.4+1.4, using data from different surveys. The author found that the spectrum at GHz frequencies is flat, inconsistent with non-thermal emission. In this work, we predict a very strong thermal component that can be important at these frequencies.
Models with spatial dependencies in massive star bow shocks \citep{2018ApJ...864...19D} show that the thermal and non-thermal emission at radio frequencies dominate in different spatial regions of the extended radio source, producing mixed spectral-index maps \citep{delvalle2025}. We would expect a similar configuration in the bubble of WR\,102. The assumption that in G2.4+1.4 the synchrotron radiation dominates near the reverse shock while free-free emission dominates in the dense shell can solve these apparently contradictory results.

We assume here spherical symmetry for modeling the bubble. However, it is noticeable in the maps presented by \citet{Prajapati2019} the asymmetry seen in the bubble structure. Optical and infrared images of the bubble also exhibit a complex structure \citep[][]{2015A&A...578A..66T}, with brighter emission toward the southeast. The radio emission also shows a strongest flux towards this region. However, strong gradients in the background emission could enhance this effect.  

The structures formed by the interaction of Wolf-Rayet winds with the medium/previous evolutionary phases can be very complex \citep[see,][among many others]{2020MNRAS.496.3906M, 2021MNRAS.507.4697M}. We have adopted a very simple model for describing the wind bubble of WR\,102, however this election is not expected to impact significantly on the results. In our radiative model, a key parameter is the termination shock velocity where particles are accelerated, even if the wind expands into a different medium the shock velocity would remain of the same order. Other two important parameters of the model are the size of the structure and the target density. The values adopted for both quantities are constrained by observations.

Our model predicts gamma-ray emission of hadronic nature. The high-energy protons interacting in the shell produce the dominant contribution. At 10\,GeV the predicted flux reaches values ca. $10^{-12}\,$erg\,s$^{-1}$\,cm$^{-2}$, which is lower than the sensitivity of the {\it Fermi-}LAT for 10 years (at $l = 0, b= 0$)\footnote{See \url{https://www.slac.stanford.edu/exp/glast/groups/canda/lat_Performance.htm}}. Our prediction is compatible with the sources reported on the LAT 14-year Source Catalog \citep[4FGL-DR4,][]{Abdollahi_2022}. The nearest point source to WR\,102 reported in the catalog is 4FGL J1744.5-2612, and both the star and its bubble are outside the confidence ellipse.

At energies $E_\gamma \sim 1\,$TeV, we predict fluxes of the order of $10^{-13}\,$erg\,s$^{-1}$\,cm$^{-3}$. With these values, the bubble of WR\,102  is an interesting target for the future Cherenkov Telescope Array Observatory (CTAO, \url{https://www.ctao.org/for-scientists/performance/}). The angular size of the bubble ($\sim 0.1^\circ$) would appear as an extended source for CTAO, making a detection more challenging. Hence, a careful detectability analysis is required.  

The gamma-ray excess from $\kappa$\,Ori \citep{2018A&A...615A..82M} supports our predictions, and could indicate that protons are also accelerated at the shock wind of massive stars.  


\section{Conclusions}\label{sec:conclusions}

In this work, we modeled the non-thermal emission expected from the stellar bubble G2.4+1.4 carved by the wind of WR\,102. Firstly, we modeled the bubble structure following the classic model for stellar bubbles, assuming spherical symmetry and used it to model the high-energy processes. We fitted the observed non-thermal emission at radio frequencies, assuming that the electrons are accelerated at the reverse shock, and emit {\it in-situ}. From this fitting, we estimated the strength of the magnetic field near the reverse shock and the fraction of the wind power in high-energy electrons. We obtained $\sim 250\,{\mu}$G and $\sim 0.03$, respectively. We also calculated the free-free radiation, competing with the synchrotron emission, and established a filling factor for the emitting volume of one half. 

We then computed the high-energy emission produced by the same population of electrons by inverse Compton and relativistic Bremsstrahlung. We also assumed that the reverse shock accelerates protons, injecting 10\% of the wind power into this high-energy particle component. The protons emit a small fraction of their energy at the reverse shock, and they propagate by diffusion towards the dense shell, emitting gamma rays there by pion decay. Additionally, we considered the contribution of cosmic rays, both electrons and protons, to the total SED when interacting with the compressed interstellar material of the bubble.\\

We conclude:
\begin{itemize}
    \item thermal emission from the ionized material competes with synchrotron at radio frequencies;
    \item our predictions are compatible with the available radio data;
    \item the confirmed non-thermal nature of the radio detection constrains the free-free emitting volume from the bubble; 
    \item the power in high-energy electrons needed to fit the observations is compatible with the available kinetic power reported for the wind;
    \item the magnetic field strength needed to fit the observations requires a magnetic flied amplification process operating at the source;
    \item the electrons reach maximum energies of the order of TeVs, while protons can reach values of the order of hundreds of TeVs and even PeVs;
    \item the protons accelerated at the reverse shock emit gamma-rays at the bubble shell, producing the dominant contribution at the high-energy domain;
    \item the PeVatron nature of single stellar bubbles would be difficult to confirm observationally, given the lower emission expected at $\sim$ 100\,TeV. This is aggravated by the softening of the particle distribution produced by diffusion;
    \item the contribution from the local cosmic rays, both leptonic and hadronic components, are negligible in this source;
    \item the bubble is transparent to gamma rays;
    \item simple estimations show that wind bubbles from WR stars in the Galaxy contribute with 4.5\% to 15\% to the total power in cosmic rays;
\end{itemize}

\begin{acknowledgments}
This study was financed by the São Paulo Research Foundation (FAPESP), Brasil (Process Number 2019/05757-9). M.V.d.V. acknowledges further support from FAPESP Process Number 2020/08729-3. L.E. acknowledges further support from FAPESP Process Number 2023/11877-2. 
\end{acknowledgments}

\software{Naima \citep{Zabalza2015}}


\bibliography{bibliography}{}
\bibliographystyle{aasjournalv7}



\end{document}